\documentclass{myfizik} 
\vol{}
\pyear{}
\received{}

\input{epsf.sty}

\def\mydate{31 January 2000}
\def\mypreprint1{UMN-TH-1839/00}
\def\mypreprint2{NUC-MINN-00/2-T}
\def\mypreprint3{BNL-NT-00/1}

\newcommand{\ben}{\begin{enumerate}}
\newcommand{\een}{\end{enumerate}}
\newcommand{\be}{\begin{equation}}
\newcommand{\ee}{\end{equation}}
\newcommand{\bse}{\begin{subequation}}
\newcommand{\ese}{\end{subequation}}
\newcommand{\bea}{\begin{eqnarray}}
\newcommand{\eea}{\end{eqnarray}}
\newcommand{\bc}{\begin{center}}
\newcommand{\ec}{\end{center}}

\def\mybig{\displaystyle \strut }
\def\myfrac#1#2{{\mybig #1\over \mybig #2}}
\def\ep{\epsilon}
\def\go{\rightarrow}
\def\myskip{\noalign{\kern 5pt}}

\author[HOSOTANI and BJORAKER]
        {
        {\bf  Yutaka HOSOTANI and 
Jefferson BJORAKER\footnote{Current address:   
{\it Brookhaven National Laboratory, Building 510A, 
Upton, NY 11973, U.S.A.}}}\\
        {\it  School of Physics and Astronomy, University of Minnesota}\\
        {\it  Minneapolis, MN 55455, U.S.A.}\\
        }
\title{Monopoles and Dyons\\ in the Pure Einstein-Yang-Mills  Theory}

\begin{document}
\maketitle

\begin{abstract}
In the pure Einstein-Yang-Mills theory in four dimensions there exist
monopole and dyon solutions.  The spectrum of the solutions is 
discrete in asymptotically flat or de Sitter space, whereas it
is continuous in asymptotically anti-de Sitter space.   The solutions are
regular everywhere and specified with their   mass, and non-Abelian
electric and magnetic charges.   In asymptotically anti-de Sitter space
a class of monopole solutions   have no node in non-Abelian magnetic
fields, and are  stable against spherically symmetric perturbations.
\end{abstract}

\section{Introduction}

In flat space there cannot be any static solution in the 
pure Yang-Mills theory in four dimensions\cite{DESER}. Only with scalar
fields monopole solutions exist, the topology of the scalar field 
playing a crucial role there.  The inclusion of the gravity opens a
possibility of having a soliton solution.  The gravitational force, being
always attractive, may balance the repulsive force of the non-Abelian
gauge fields.  Such configurations were indeed found in
asymptotically flat and de Sitter space some time ago 
\cite{BARTNIK}-\cite{VOLKOV2}.
Unfortunately all of them turned out unstable against small
perturbations \cite{ZHOU}-\cite{Kanti}.

The situation drastically changes in asymptotically anti-de Sitter
space.  The negative cosmological constant ($\Lambda <0$) provides
negative pressure and energy density,  making a class of monopole
and dyon configurations stable \cite{WINSTANLEY}-\cite{Bjoraker2}.  We
review the current status of such solutions.

\section{Equations}

The equations of motion in the Einstein-Yang-Mills theory are
\bea
&&\hskip -1.cm
R^{\mu\nu} - {1\over 2}g^{\mu\nu} (R - 2 \Lambda) = 8\pi G ~
T^{\mu\nu}\cr
\noalign{\kern 5pt}
&&\hskip -1.cm
{F^{\mu\nu}}_{;\mu} + e [A_\mu, F^{\mu\nu} ] = 0 ~~.
\label{EYM-eq}
\eea
To find soliton solutions we make a spherically symmetric, static
ansatz:
\bea
&&\hskip -1.cm
ds^2 = -\myfrac{H(r)}{p(r)^2} \, dt^2 + \myfrac{dr^2}{H(r)} 
 + r^2 (d\theta^2 + \sin^2 \theta \, d\phi^2) \cr
\noalign{\kern 5pt}
&&\hskip -1.cm
A = {\tau^j\over 2e} \bigg\{ 
u(r) \,  {x_j\over r} \, dt 
-   {1-w(r) \over r^2}\, \ep_{jkl} x_k \,dx_l \bigg\} 
\label{ansatz1}
\eea
with boundary conditions $u(0)=0$ and $H(0)=p(0)=w(0)=1$.  It is  
convenient to parameterize
\be 
H(r) = 1 - {2m(r)\over r} - {\Lambda r^2\over 3} ~~,
\label{m-function}
\ee
where $m(r)$ is the mass contained inside the radius $r$.  $p(r)=1$
and $m(r)=0$ corresponds to the Minkowski, de Sitter, or
anti-de Sitter space for $\Lambda=0$, $>0$, or $<0$, respectively.
$u(r)$ and $w(r)$ represent the electric and magnetic Yang-Mills fields,
respectively.  

The equations in (\ref{EYM-eq}) reduce to 
\bea
&&\hskip -1.cm
\big( r^2 p u' \big)' = {2p\over H} \, uw^2 \cr
\myskip
&&\hskip -1.cm
\Big( {H\over p }\, w' \Big)'
= {w(w^2-1)\over r^2 p} - {p\over H} \, u^2 w \cr
\myskip
&&\hskip -1.cm
m' = {4\pi G\over e^2} \, 
\bigg\{ {1\over 2} r^2 p^2 u'^2 + {(1-w^2)^2\over 2r^2}
+ {p^2 \over H}\, u^2 w^2 + H w'^2\bigg\}  \cr
\myskip
&&\hskip -1.cm
p'~ = - {8\pi G\over e^2} \, {p\over rH} \, 
\bigg\{ {p^2 \over H}\, u^2 w^2 + H w'^2 \bigg\} ~.
\label{EYM-eq2}
\eea
Near the origin $u= ar$, $w= 1 - br^2$, 
$m= v(a^2 + 4b^2) r^3/2$, and $p= 1 - v(a^2 + 4b^2) r^2$,
where $(a,b)$ are two parameters to be fixed and 
$v=4\pi G/e^2$.  With given $(a,b)$
the equations in (\ref{EYM-eq2}) are numerically integrated from
$r=0$ to $r=\infty$.

In general, solutions blow up at finite $r$, unless $(a,b)$ take
special values.   We are looking for everywhere regular soliton
configurations with a finite ADM mass $M = m(\infty)$.  

\section{Conserved charges}

Non-Abelian solitons are characterized by the ADM mass and non-Abelian
electric and magnetic charges.  The kinematical identities
$({F^{\mu\nu}}_{;\mu})_{;\nu}=0$ and 
$({ \tilde {F}^{\mu\nu}\, }_{;\mu} )_{;\nu}=0$ lead to
conserved charges given by
$\int dS_k \, \sqrt{-g} \, (F^{k0} , \tilde F^{k0})$. They are
gauge variant, and therefore there are infinitely many conserved
charges.  Most of them vanish for solutions under consideration.
Non-vanishing charges are
\bea
\pmatrix{Q_E\cr Q_M} &=& {e\over 4\pi} \int dS_k \,
 \sqrt{-g} \, {\rm Tr~} \tau_r \pmatrix{F^{k0}\cr \tilde F^{k0}} ~~, 
\hskip .5cm \tau_r = {x^j \tau^j\over r} \cr
\myskip
&=& \pmatrix{u_1 p_0\cr 1 - w_0^2} ~~.
\label{charge1}
\eea
In the second equality the coefficients $u_1$, $p_0$, and $w_0$ are
defined by the asymptotic expansion 
$u \sim u_0 + (u_1/r) + \cdots$ etc..
These two charges are conserved as there exists a unitary matrix $S$
satisfying $\tau_r = S \tau_3 S^{-1}$.

\section{Solutions in the $\Lambda=0$ or $\Lambda>0$ case}

It has been shown in \cite{GALTSOV} and \cite{Bjoraker2} that
solutions are electrically neutral ($a= 0$, $u(r)= 0$).  Solutions
exist only for a discrete set of values of $b$.  In the $\Lambda=0$ case,
$w_0=w(\infty)=\pm 1$ so that $Q_M=0$.  In the $\Lambda>0$ case,  $Q_M
\not= 0$. The Bartnik-McKinnon solution in the $\Lambda=0$ case is
depicted in fig.\ 1.

\begin{figure}[tbh]\centering
 \leavevmode 
\mbox{
\epsfxsize=9.0cm \epsfbox{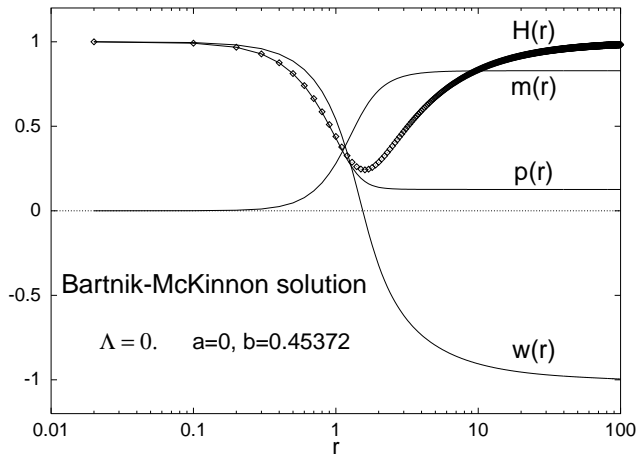}}
\caption{Bartnik-McKinnon solution in the $\Lambda=0$ theory.}
\end{figure}

Solutions are characterized by the number of nodes, $n$, in $w(r)$;
$n=1,2,3, \cdots.$
All of these solutions are shown to be unstable against small
spherically symmetric perturbations \cite{ZHOU}-\cite{VOLKOV3}.

\section{Solutions in the $\Lambda < 0$ case}

\subsection{Configurations}

The $\Lambda < 0$ case is qualitatively different in many respects
from the $\Lambda=0$ or $\Lambda > 0$ case.  
First there are dyonic solutions in which electric fields are
non-vanishing; $u(r) \not= 0$.  Secondly solutions exist
in a finite continuum region in the parameter space $(a,b)$
as opposed to discrete points.
Thirdly there exist solutions with no node ($n=0$) in $w(r)$.  

A typical monopole solution with no node in $w$ is depicted in 
fig.\ 2.  Depending on the value of $b$, the asymptotic value 
$w(\infty)$ can be either greater than 1, or between 0 and 1, 
or negative.  

\begin{figure}[thb]\centering
 \leavevmode 
\mbox{
\epsfxsize=9.0cm \epsfbox{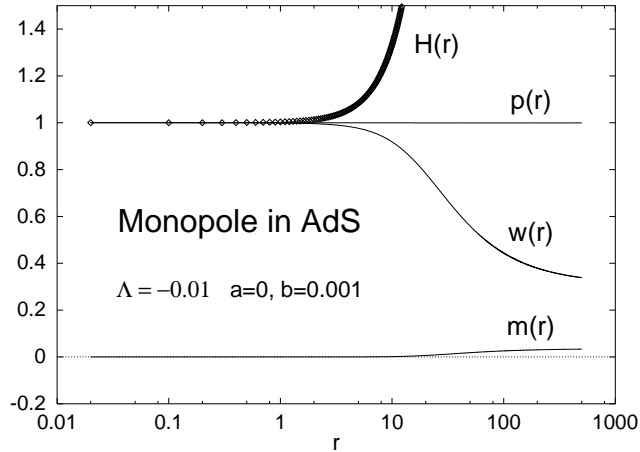}}
\caption{Monopole solution in the $\Lambda < 0$ theory. There are a
continuum of solutions. Dyon solutions have similar behavior, with
the additional $u(r)$ monotonically increasing from zero to
the asymptotic value in the range $(0 \sim 0.2)$.}
\end{figure}

\subsection{Monopole and dyon spectrum}

When $a=0$ and $b$ is varied, a continuum of monopole solutions 
are generated.  With $\Lambda$ given, solutions appear in a finite
number of  branches.  The number of branches increases as $\Lambda \go 0$.
For $\Lambda=-0.01$ there are only two branches, which are displayed in
fig.\ 3.

The upper branch ends near the point $(Q_M=1, M=1)$.  The end point
corresponds to the critical spacetime geometry discussed in the next
subsection.

Solutions with no node in $w(r)$ are special. They are stable
against small perturbations as discussed in Section 6.  They exist in a
limited region in the parameter space as depicted in fig.\ 4.

\begin{figure}[tbh]
\centering \leavevmode 
\mbox{
\epsfxsize=8.cm \epsfbox{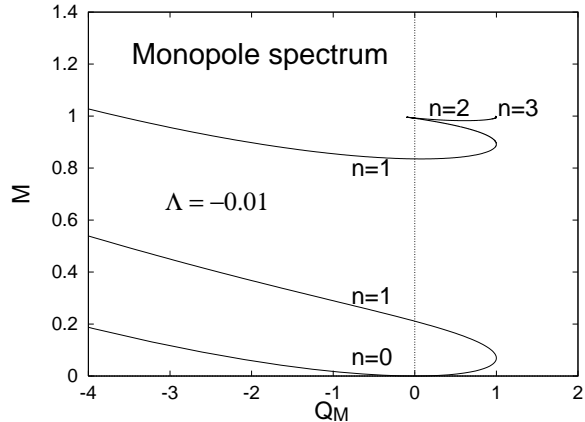}}
\caption{Monopole spectrum at $\Lambda=-0.01$. $n$ is the number of
the nodes in $w(r)$.}
\end{figure}
\begin{figure}[bth]
\centering \leavevmode 
\mbox{
\epsfxsize=8cm \epsfbox{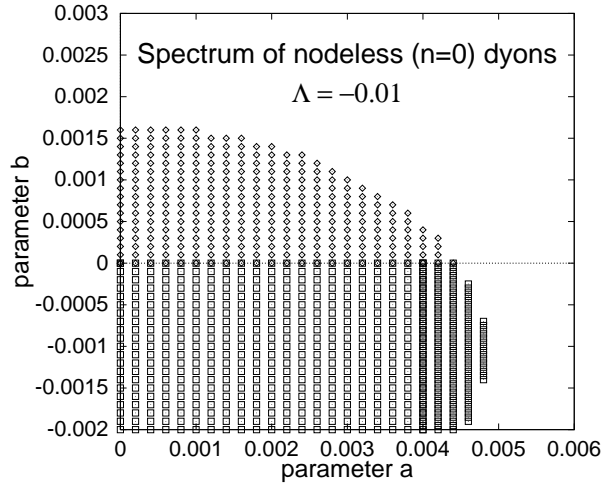}}
\caption{Spectrum of nodeless dyons.}
\end{figure}
\vskip 1cm

\subsection{Critical spacetime}
When the parameter $b$ is increased, the solution either blows up
or reaches a critical configuration.  The end point in the upper branch 
in fig. 3 represents such a critical spacetime with $b=b_c=0.70$.
$H(r)$ vanishes at $r=r_h$.  However, $r=r_h$ is not a standard
event horizon appearing in black hole solutions.

When $b$ is very close to $b_c$, $H(r)$ almost vanishes at $r\sim r_h$.
One of such configurations is displayed in fig.\ 5.

\begin{figure}[bth]\centering
 \leavevmode 
\mbox{
\epsfxsize=8.cm \epsfbox{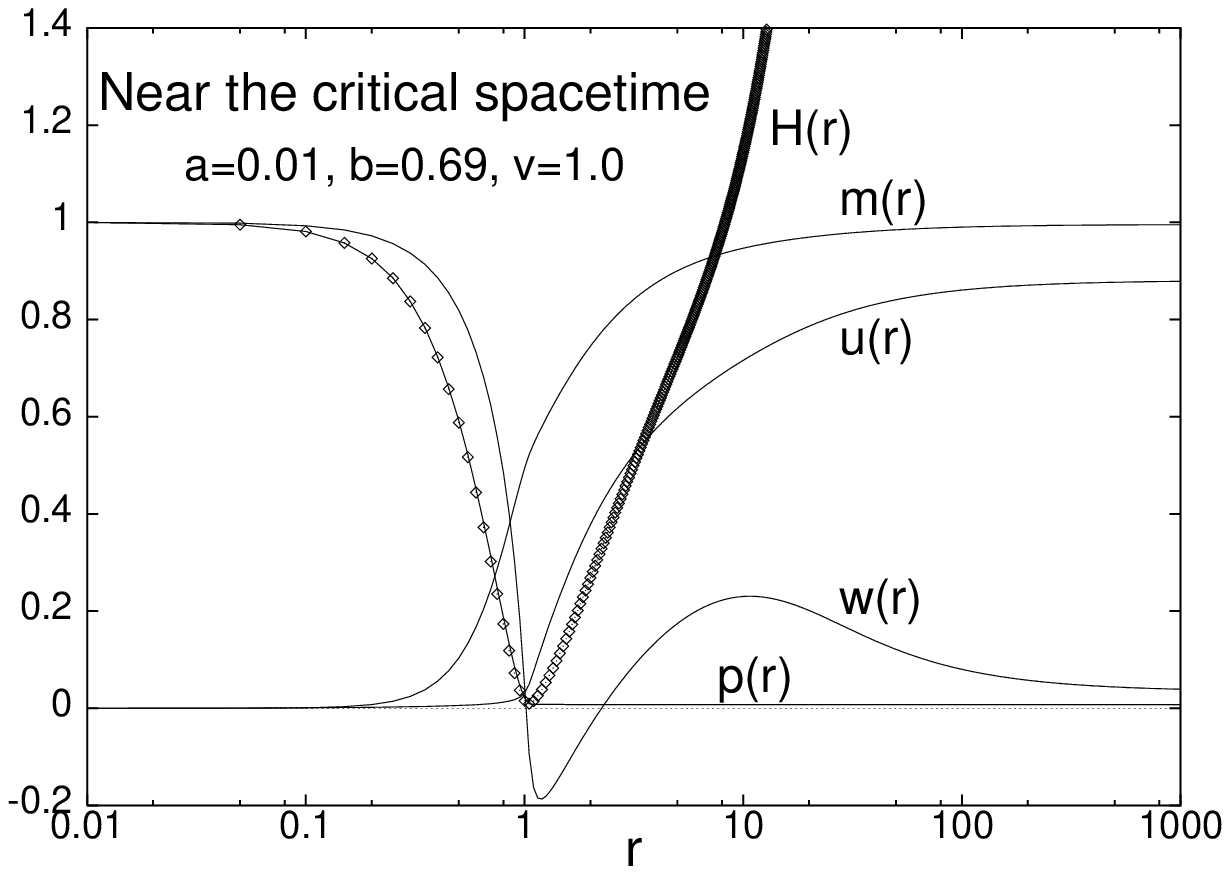}}
\caption{Dyon solution very close to the critical spacetime. At the
critical value $b_c$ the space ends at $r_h$.  $p(r)=0$ for 
$r \ge r_h$.}
\end{figure}

At $b=b_c$, $H(r)$ becomes tangent to the axis at $r_h$.  Further 
$p(r)$ vanishes.  In other words the space ends at $r=r_h$.
It is an open question whether such configurations really represent
possible spacetime.   

These critical spacetimes have universal behavior.  Their magnetic
charge is quantized, $Q_M =1$, whereas their electric charge
$Q_E$ is not.  There are two additional parameters, $\Lambda$ and
$v=4\pi G/e^2$.  When $v|\Lambda | \ll 1$, the critical spacetime
is described near $r_h$ by
\bea
r_h ~~ &=& {1\over 2|\Lambda|} \Big( \sqrt{1+4v|\Lambda|} - 1 \Big) 
   \sim \sqrt{v} \cr
\myskip
w(r) &\sim& 2 y^{1/2} \cr
\myskip
H(r) &\sim& 4 y^2 \cr
\myskip
m(r) &\sim& {1\over 2}\sqrt{v} (1 - y) \cr
\myskip
p(r) &\sim& p_0 y^2 \cr
\myskip
&&\hskip -.5cm \hbox{where}  \hskip .5cm 
   y = 1 - {r\over r_h} \ge 0~~.
\label{universality}
\eea
Except for $p_0$ all coefficients and critical exponents are
independent of $Q_E$ and $|\Lambda| (\ll 1/v)$.

\section{Stability}

The stability of the solutions is examined by considering small
perturbations.  If they exponentially grow in time, the solutions
are unstable, whereas if they remain small, they are stable.
In the linearized theory, the problem is reduced to finding
eigenvalues of a Schr\"odinger equation.  

The analysis is simplified in the tortoise coordinate $\rho$
defined by $d\rho/dr = p/H$.  The range of $\rho$ is finite
for $\Lambda < 0$ ; $0 \le \rho \le \rho_{\rm max}$.
For monopole solutions
\bea
&&\hskip 0.cm
\bigg\{ - {d^2\over d\rho^2} + U(\rho) \bigg\}  \chi
    = \omega^2 \chi   \label{Schrodinger} \\
\noalign{\kern 12pt}
&&\hskip -1cm
\hbox{(i)  Odd parity perturbations} \cr
\noalign{\kern 8pt}
&&\hskip -.5cm
U_{\rm odd} = \frac{H}{r^2p^2}(1+w^2)
+\frac{2}{w^2}\left(\frac{dw}{d\rho}\right)^2 ~~,  
\label{U_odd} \\
\noalign{\kern 8pt}
&&\hskip -0.5cm
\delta\nu = {w\over r^2 p} \, \chi_{\rm odd}~~~,~~~
\delta \tilde w = - {1\over 2w} \, {d\over d\rho} (r^2 p \delta \nu) \cr
\noalign{\kern 8pt}
&&\hskip -1cm
\hbox{(ii)  Even parity perturbations} \cr
\noalign{\kern 8pt}
&&\hskip -.5cm
U_{\rm even} = \frac{H }{r^2 p^2} \, (3w^2-1)
+ 4 v \frac{d}{d\rho}\bigg( \frac{Hw'^2}{pr}  \bigg) 
\label{U_even} \\
\noalign{\kern 8pt}
&&\hskip -0.5cm
\delta w = \chi_{\rm even} ~~,~~ 
{\delta H\over H} = - {4v\over r} \, w'\delta w ~~,~~
\Big( {\delta p\over p} \Big)' = - {4v\over r} \, w'\delta w' \cr
\nonumber
\eea
Here  $\, ' \,$ denotes a $r$-derivative.  The boundary condition for
odd parity perturbations is given by  $\chi_{\rm odd}=0$ at $\rho=0$ and
$d(w \chi_{\rm odd})/d\rho =0$ at $\rho=\rho_{\rm max}$.
For even parity perturbations $\chi_{\rm even}=0$ at both ends.
If Eq.\ (\ref{Schrodinger}) admits no boundstate ($\omega^2 < 0$), 
then the solution is stable.

Although $U_{\rm odd}(\rho)$ is positive definite,  the corresponding
eigenvalues may not be positive due to the nontrivial boundary condition
imposed on $\chi_{\rm odd}$.  For solutions with no node in $w(r)$,
both $U_{\rm odd}$ and $U_{\rm even}$ behave as $2/\rho^2$ near
the origin, but are regular elsewhere.  One can prove that
all $\omega^2_{\rm odd}, \omega^2_{\rm even} > 0$.  In other words
nodeless monopole solutions are stable against spherically
symmetric perturbations.

When $w$ has $n$ nodes at $\rho_k$ ($k=1, \cdots, n$), 
$U_{\rm odd}$ become singular there.  There appear $n$ negative
$\omega^2_{\rm odd}$ modes, which generally diverges at the 
singularities.  $U_{\rm even}$ also becomes negative in the vicinity of
the  nodes, and there appear negative $\omega^2_{\rm even}$.  Solutions
with nodes in $w(r)$ are unstable.

\section{The $\Lambda \go 0$ limit}

When the cosmological constant approaches zero, more and more 
branches of monopole solutions emerge.  In the $\Lambda \go 0$ limit
the spectrum becomes discrete,  there appearing infinitely many 
(unstable) solutions.

How is it possible? The nodeless, stable solutions must disappear.
One parameter family of solutions must collapse into one point
in the moduli space of solutions.  In fig.\ 6 we have plotted
the monopole spectrum with various values of $\Lambda$.  One can
see that as $\Lambda$ approaches zero, new branches of solutions 
appear, and each branch collapses to a point in the $\Lambda\go 0$
limit.   The nodeless solutions disappear as their ADM mass vanishes.

The result is indicative of  a fractal structure in the moduli
space of the solutions.

\begin{figure}[h]
\centering \leavevmode 
\mbox{
\epsfxsize=8.cm \epsfbox{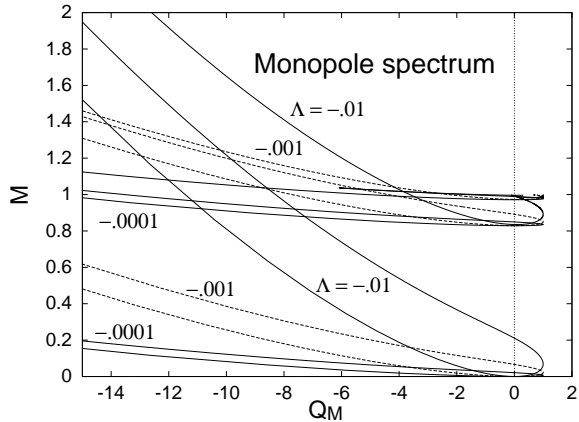}}
\caption{Monopole spectrum with varying $\Lambda$.}
\end{figure}

\section{Summary}

We have shown that there exist stable monopole and dyon 
solutions in the Einstein-Yang-Mills theory in asymptotically
anti-de Sitter space.  They have a continuous spectrum.
Their implication to physics is yet to be examined.

\vskip 1cm

\leftline{\bf Acknowledgments}

This work was supported in part    by the U.S.\ Department of
Energy under contracts DE-FG02-94ER-40823 and DE-FG02-87ER40328.

\end{document}